\documentclass[useAMS,usenatbib]{mn2e}
\usepackage{psfig}
\usepackage{epsfig}
\usepackage{amsbsy} 
\usepackage{amssymb}

\newcommand{\be}{\begin{equation}}
\newcommand{\ee}{\end{equation}}
\newcommand{\bea}{\begin{eqnarray}}
\newcommand{\eea}{\end{eqnarray}}

\newcommand{\n}{\label}

\newcommand{\COSMOMC}{{\sc cosmomc}}
\newcommand{\CAMB}{{\sc camb}}

\def\apj{{  ApJ}}
\def\apjs{{  ApJ S.}}
\def\prd{{  Phys. Rev.}{D}}

\def\mnras{{MNRAS}}

\def\lt{<}
\def\gt{>}
\def\amp{\&~}
\newcommand{\lsim}{\lesssim}
\newcommand{\gsim}{\gtrsim}
\def \lleq {\lower0.9ex\hbox{$\buildrel \lt \over \sim$} ~}

\title[Constraining dark energy with galaxy clusters, supernovae and
the CMB]{Constraining Dark Energy with X-ray Galaxy Clusters,
Supernovae and the Cosmic Microwave Background} \author[D.~Rapetti,
S.~W.~Allen and J.~Weller] {David Rapetti${}^{1,2,3}$\thanks{Email:
drapetti@ast.cam.ac.uk}, Steven W.~Allen${}^{1,3}$ and Jochen
Weller${}^{1,4,5}$ \\ ${}^1$ Institute of Astronomy, University of
Cambridge, Madingley Road, Cambridge CB3 0HA, UK\\ ${}^2$ Departament
d'Astronomia i Meteorologia, Universitat de Barcelona, Mart\'{\i} i
Franqu\`es 1, 08028 Barcelona, Spain. \\ ${}^3$ Kavli Institute for
Particle Astrophysics and Cosmology, Stanford University, 382 Via
Pueblo Mall, Stanford 94305-4060, USA. \\${}^4$ NASA/Fermilab
Astrophysics Group, Fermi National Accelerator Laboratory, Batavia, IL
60510-0500, USA. \\${}^5$ Department of Physics and Astronomy,
University College London, Gower Street, London WC1E 6BT, UK.}

\begin{document}
\date{Accepted ???, Received ???; in original form \today}
\pagerange{\pageref{firstpage}--\pageref{lastpage}} \pubyear{2004}

\maketitle
\label{firstpage}

\begin{abstract}
  We present new constraints on the evolution of dark energy from an
  analysis of Cosmic Microwave Background, supernova and X-ray galaxy
  cluster data. Our analysis employs a minimum of priors and exploits
  the complementary nature of these data sets. We examine a series of
  dark energy models with up to three free parameters: the current
  dark energy equation of state $w_{\rm 0}$, the early time equation
  of state $w_{\rm et}$ and the scale factor at transition, $a_{\rm
  t}$.  From a combined analysis of all three data sets, assuming a
  constant equation of state and that the Universe is flat, we measure
  $w_{\rm 0}=-1.05^{+0.10}_{-0.12}$. Including $w_{\rm et}$ as a free
  parameter and allowing the transition scale factor to vary over the
  range $0.5\lt a_{\rm t}\lt 0.95$ where the data sets have
  discriminating power, we measure $w_{\rm 0} =-1.27^{+0.33}_{-0.39}$
  and $w_{\rm et} =-0.66^{+0.44}_{-0.62}$. We find no significant
  evidence for evolution in the dark energy equation of state
  parameter with redshift. Marginal hints of evolution in the
  supernovae data become less significant when the cluster constraints
  are also included in the analysis. The complementary nature of the
  data sets leads to a tight constraint on the mean matter density,
  $\Omega_{\rm m}$ and alleviates a number of other parameter
  degeneracies, including that between the scalar spectral index
  $n_{\rm s}$, the physical baryon density $\Omega_{\rm b}h^2$ and the
  optical depth $\tau$. This complementary nature also allows us to
  examine models in which we drop the prior on the curvature. For
  non-flat models with a constant equation of state, we measure
  $w_{\rm 0}=-1.09^{+0.12}_{-0.15}$ and obtain a tight constraint on
  the current dark energy density, $\Omega_{\rm de}=0.70\pm0.03$. For
  dark energy models other than a cosmological constant,
  energy--momentum conservation requires the inclusion of spatial
  perturbations in the dark energy component. Our analysis includes
  such perturbations, assuming a sound speed $c_{\rm s}^2 =1$ in the
  dark energy fluid as expected for Quintessence scenarios. For our
  most general dark energy model, not including such perturbations
  would lead to spurious constraints on $w_{\rm et}$ which would be
  tighter than those mentioned above by approximately a factor two
  with the current data.
\end{abstract}


\section{Introduction}
The precise measurement of the Cosmic Microwave Background (CMB) made
with the Wilkinson Microwave Anisotropy Probe (WMAP)
\citep{Hinshaw03,Kogut03} has improved our knowledge of a wide range
of cosmological parameters.  However, a number of degeneracies between
parameters exist which cannot be broken with current CMB data alone
and which require the introduction of other, complementary data sets.

Some of the most important parameters and degeneracies concern dark
energy and its equation of state. Since observations of distant type
Ia supernovae (SNIa) first indicated that the expansion of the
Universe is accelerating \citep{Riess:98,Perlmutter98}, there has been
enormous interest in this topic. The most straightforward way to
incorporate an accelerated expansion into cosmological models is by
adding a constant term to the Einstein equations - the cosmological
constant. However, this leads to an extreme fine tuning problem
wherein one must adjust the initial density of this constant to
$10^{-120} M_{\rm pl}^4$ in natural Planck units. To alleviate this, a
scalar field model, dubbed Quintessence, was introduced which, when
the potential is carefully chosen, can avoid the fine tuning of
initial conditions
\citep{Peebles:88,Ratra:88,Wetterich:88,Ferreira:98,Caldwell:98,Zlatev:98}.
When describing the background evolution of the Universe with such
models, it is sufficient to know the equation of state for the dark
energy i.e. the ratio of pressure and energy density, $w = p_{\rm
de}/\rho_{\rm de}$. Whilst a cosmological constant has $w=-1$ at all
times, for most dark energy models the equation of state parameter is
an evolving function of redshift, $w=w(z)$.

In order to learn more about the origin of cosmic acceleration and
dark energy, it is crucial to constrain the evolution of the dark
energy equation of state. In the first case, this requires us to
examine whether the accelerated expansion can be described by a
cosmological constant or if there is need to go beyond this
description.  Most early attempts to parameterise the evolution of
dark energy were carried out as feasibility studies for future
supernovae experiments
\citep{Huterer:98,Efstathiou:99,Saini:99,Maor:00,Astier:00,Weller:00,Weller01}.
However, recent improvements in the data for high-redshift SNIa and
the arrival of other, complementary constraints means that we can now
start to ask the same questions of real data \citep{Knop:03,Riess:04}.

The data for SNIa can be used to measure the luminosity distances to
these sources independent of their redshifts. This constrains a
combination of the dark matter and dark energy densities in a
different way to observations of CMB anisotropies. The combination of
the two data sets is therefore useful in breaking parameter
degeneracies. However, the simplest, linear expansion in redshift for
the dark energy equation of state advocated by e.g.  \cite{Maor:00},
\cite{Astier:00} and \cite{Weller:00,Weller01} cannot be applied to
the high redshifts probed by the CMB. For this reason, the linear
parameterisation was extended by \cite{Chevallier:01} and
\cite{Linder:03} to a model in which the equation of state at low
redshifts (late times) $w_{\rm 0}$, and at high redshifts (early
times) $w_{\rm et}$, could be specified separately. Although more
suitable than the low redshift linear expansion for the analysis of
CMB data, this parameterisation also has a short-coming in that the
transition between $w_{\rm 0}$ and $w_{\rm et}$ always occurs at
redshift $z=1$ and with a fixed transition rate, which is not
representative of the full range of scalar field dark energy models of
interest (see e.~g.~\cite{Weller01}).  \cite{Corasaniti:03} and
\cite{Bassett:04} extended this prescription further, allowing the
transition to occur at an arbitrary time and
rate. \cite{Corasaniti:04a} applied this extended model to a
combination of SNIa, CMB and galaxy redshift survey data, noting the
presence of strong degeneracies between a number of the derived
parameters.  The analysis of \cite{Corasaniti:04a} included a limited
exploration of models with an equation of state $w\lt -1$: so called
phantom models \citep{Caldwell:02}. While the extension to $w\lt -1$
is challenging in terms of the physics involved \citep{Carroll:03}, it
is interesting from a phenomenological point of view, particularly
given that the best-fit to current supernovae data is obtained for
models with $w\lt -1$ \citep{Riess:04}.

It is important to note that the analysis of \cite{Corasaniti:04a} did
not include dark energy perturbations for models crossing $w=-1$.
While a cosmological constant is spatially homogeneous, this is not
true for an arbitrary dark energy fluid or Quintessence. One must
include perturbations in the dark energy component, not just for
consistency reasons but also because the exclusion of them can lead to
erroneously tight constraints on $w$ from large-scale CMB anisotropies
\citep{Weller:03}.

It was realised by \cite{Maor:00} and \cite{Weller:00,Weller01} that
in order to constrain the evolution of the equation of state with
supernovae observations, it is necessary to use a tight prior on the
mean matter density of the Universe, $\Omega_{\rm m}$. Recent
measurements of the gas fraction in X-ray luminous, dynamically
relaxed clusters made with the Chandra X-ray Observatory provide one
of our best constraints on $\Omega_{\rm m}$ \citep{Allen:04}. These
data also provide a direct and independent method by which to measure
the acceleration of the Universe, providing additional discriminating
power for dark energy studies. As we shall demonstrate here, the
combination of CMB and X-ray cluster data can also play an important
role in breaking other key parameter degeneracies (see also
\cite{Allen:03}).  For these reasons, we have used a combination of
X-ray gas fraction, CMB and SNIa data in this study.

In the following sections we first introduce our choice of
parameterisations for the dark energy equation of state. We then
discuss the individual data sets and how they probe cosmology. Our
results are presented in Section \ref{constraints}. Section
\ref{discussion} discusses the results and summarises our conclusions.

\section{Dark Energy Model}
\n{eos}

A number of different parameterisations for the evolution of the dark
energy equation of state parameter, $w(z)$, have been discussed in the
literature. The simplest is the linear parameterisation: $w(z)=w_{\rm
0}+w'z$ \citep{Maor:00,Weller:00,Weller01,Astier:00}. However, as
mentioned above, this model is not compatible with CMB data since it
diverges at high redshift. \cite{Chevallier:01} (see also
\cite{Linder:03}) proposed an extended parameterisation which avoids
this problem, with $w(z)=w_{\rm 0}+w_{1}z/(1+z)$. This model can in
principle be used to distinguish a cosmological constant from other
forms of dark energy with a varying $w$. However, this
parameterisation is not representative of standard Quintessence
models. \cite{Corasaniti:03} proposed a generalised parameterisation
which is better suited to the problem. However, that model includes
four parameters and exhibits large degeneracies when applied current
data.

Here, we use an extension of the model discussed by
\cite{Chevallier:01} and \cite{Linder:03}, which stops short of the
full extension suggested by \cite{Corasaniti:03}. The primary
short-coming of the parameterisation proposed by \cite{Chevallier:01,
Linder:03} is that it uses a fixed redshift, $z=1$, for the transition
between the current value of the equation of state and the value at
early times, $w_{\rm et}=w_{\rm 0}+w_{1}$. Our model introduces one
extra parameter, $z_{\rm t}$, the transition redshift between $w_{\rm
et}$ and $w_{\rm 0}$, such that \be w=\frac{w_{\rm et}z+w_{\rm
0}z_{\rm t}}{z+z_{\rm t}}=\frac{w_{\rm et}(1-a)a_{\rm t}+w_{\rm
0}(1-a_{\rm t})a} {a(1-2a_{\rm t})+a_{\rm t}}\; , \n{eqn:wa} \ee where
$a_{\rm t}$ is the transition scale factor. (The parameterisation of
\cite{Corasaniti:03} also introduces an arbitrary transition rate
between $w_{\rm 0}$ and $w_{\rm et}$.)

Energy conservation of the dark energy fluid results in evolution
of the energy density with the scale factor, such that 
\be
\rho_{\rm de}(a)=\rho_{{\rm de,}0}a^{-3}e^{-3\int_{1}^{a}{\frac{w(a')}{a'}da'}},\; 
\n{eqn:conservation}
\ee
where $\rho_{{\rm de,}0}$ is the energy density of the dark 
energy fluid today. Using the parameterisation of equation
(\ref{eqn:wa}) we obtain
\begin{equation}
\int_{1}^{a}{\frac{w(a')}{a'}da'}=w_{\rm et}\ln a + 
(w_{\rm et}-w_{\rm 0})g(a;a_{\rm t})\; ,
\n{eqn:intwa}
\end{equation}
with
\begin{equation}
g(a;a_{\rm t})=\left(\frac{1-a_{\rm t}}{1-2a_{\rm t}}\right)\ln \left( \frac{ a(1-a_{\rm t}) }{a(1-2a_{\rm t})+a_{\rm t}} \right)\;.
\n{intwa2}
\end{equation}

Setting $z_{\rm t} = 1$ or $a_{\rm t} = 1/2$, we recover the parameterisation
of \cite{Linder:03}. Hereafter, 
we shall refer to this as the $z_{\rm t}=1$ dark energy model.
From the Friedmann equation, the evolution of the 
Hubble parameter $H(z)=H_0 E(z)$ is given by
\be
E(z) = \sqrt{ \Omega_{\rm m} (1+z)^3 + \Omega_{\rm de} f(z) + \Omega_{\rm k} (1+z)^2}\;, 
\n{hubble}
\ee
with
\be
f(z) = (1+z)^{3(1+w_{\rm et})} e^{-3(w_{\rm et}-w_{\rm 0})g(z;z_{\rm t})},
\n{f}
\ee
where $\Omega_{\rm m}$, $\Omega_{\rm de}$, $\Omega_{\rm k}$ are the
matter, dark energy and curvature densities in units of the critical density.

\section{Data Analysis}
\n{analysis}

We have performed a likelihood analysis using three cosmological data
sets: CMB, SNIa and the X-ray cluster gas fraction.

For the CMB analysis we have modified the
\CAMB\footnote{http://camb.info} code~\citep{Lewis99} to include the
relevant dark energy equation of state parameters. For the calculation
of CMB spectra, we have accounted for the effects perturbations in the
dark energy component \citep{Weller:03}. We assume that the sound
speed of the dark energy fluid, $c_{\rm s}^2 =1$, a choice that is
well motivated for standard Quintessence scenarios \citep{Weller:03},
although some other well-motivated dark energy models such as
k--essence scenarios \citep{armendariz} include an evolving sound
speed. We note the presence of an extra term in the perturbation
equations due to the variation of the equation of state with time,
which sources the density perturbation with the velocity
perturbation. This effect will be discussed in a forthcoming
publication \citep{Rapetti:05}.

We use three CMB data sets: WMAP \citep{Verde03,Hinshaw03,Kogut03}
(including the temperature-polarization cross-correlation data), the
Cosmic Background Imager (CBI) \citep{Pearson03} and the Arcminute
Cosmology Bolometer Array Receiver (ACBAR) \citep{Kuo04}. The latter
data sets provide important information on smaller scales
($\ell\gt800$).

For the SNIa analysis, we use the gold sample of \cite{Riess:04},
marginalising analytically over the absolute magnitude $M$ as a
``nuisance parameter''. We fit the extinction-corrected distance
moduli, $\mu_0 = m-M=5\log d_L +25 $, where $m$ is the apparent
magnitude and $d_L$ is the luminosity distance in units of Mpc defined
as
\begin{equation}
d_L= \frac{c (1+z)}{H_0 \sqrt{\Omega_k }}{\rm sinh}\left(
\sqrt{\Omega_k }\int_0^z\frac{dz}{\sqrt{E(z)}}\right)\; ,
\n{ludist}
\end{equation}
where $\Omega_{\rm k}=1-\Omega_{\rm m}-\Omega_{\rm de}$.

\begin{figure}
\includegraphics[width=3.2in]{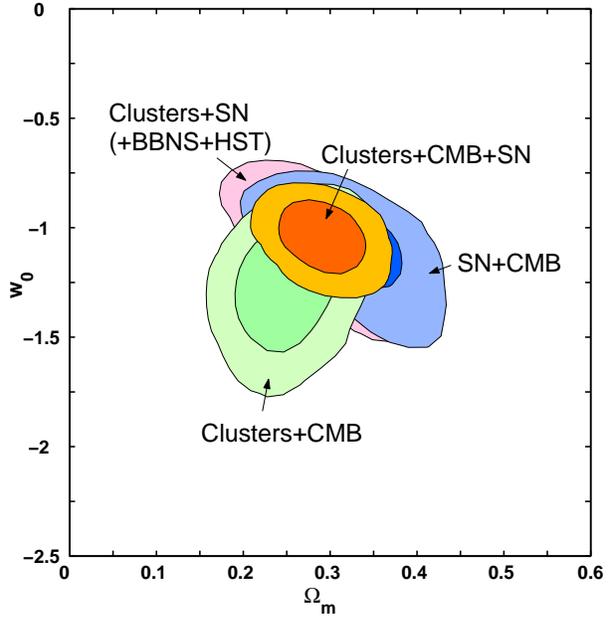}
\caption{The 68.3 and 95.4 per cent confidence limits in the
    $(\Omega_{\rm m},w_{\rm 0})$ plane for the various pairs of data
    sets and for all three data sets combined. A constant dark energy
    equation of state parameter is assumed. }
\label{wconst}
\end{figure} 

For the X-ray gas mass fraction analysis, we use the data and method
of \cite{Allen:04}, fitting the apparent redshift evolution of the
cluster gas fraction with the model
\begin{equation}
f_{\rm gas}^{\rm SCDM}(z) = \frac{ b\, \Omega_{\rm b}} {\left(1+0.19
\sqrt{h}\right) \Omega_{\rm m}} \left[ \frac{d_{\rm
A}^{\rm SCDM}(z)}{d_{\rm A}^{\rm de}(z)} \right]^{1.5}\;,
\label{eq:fgas}
\end{equation}
where $d_{\rm A}^{\rm de}(z)$ and $d_{\rm A}^{\rm SCDM}(z)$ are the
angular diameter distances ($d_A=d_L/(1+z)^2$) to the clusters for a
given dark energy (de) model and the reference standard cold dark
matter cosmology, respectively. $\Omega_{\rm b}$ is the mean baryonic
matter density of the Universe in units of the critical density, $H_0
= 100\,h\,{\rm km}\,{\rm sec}^{-1}\, {\rm Mpc}^{-1}$ and $b$ is a bias
factor that accounts for the (relatively small amount of) baryonic
material expelled from galaxy clusters as they form.  Following
\cite{Allen:04}, we adopt a Gaussian prior on $b=0.824\pm0.089$, which
is appropriate for clusters of the masses studied here. Note that the
prior on $b$ includes a 10 per cent allowance for systematic
uncertainties in the normalisation of the $f_{\rm gas}(z)$ curve,
although we note that even doubling this systematic uncertainty has
only a small effect on the results \citep{Allen:04}.

\begin{figure*}
\includegraphics[width=3.2in]{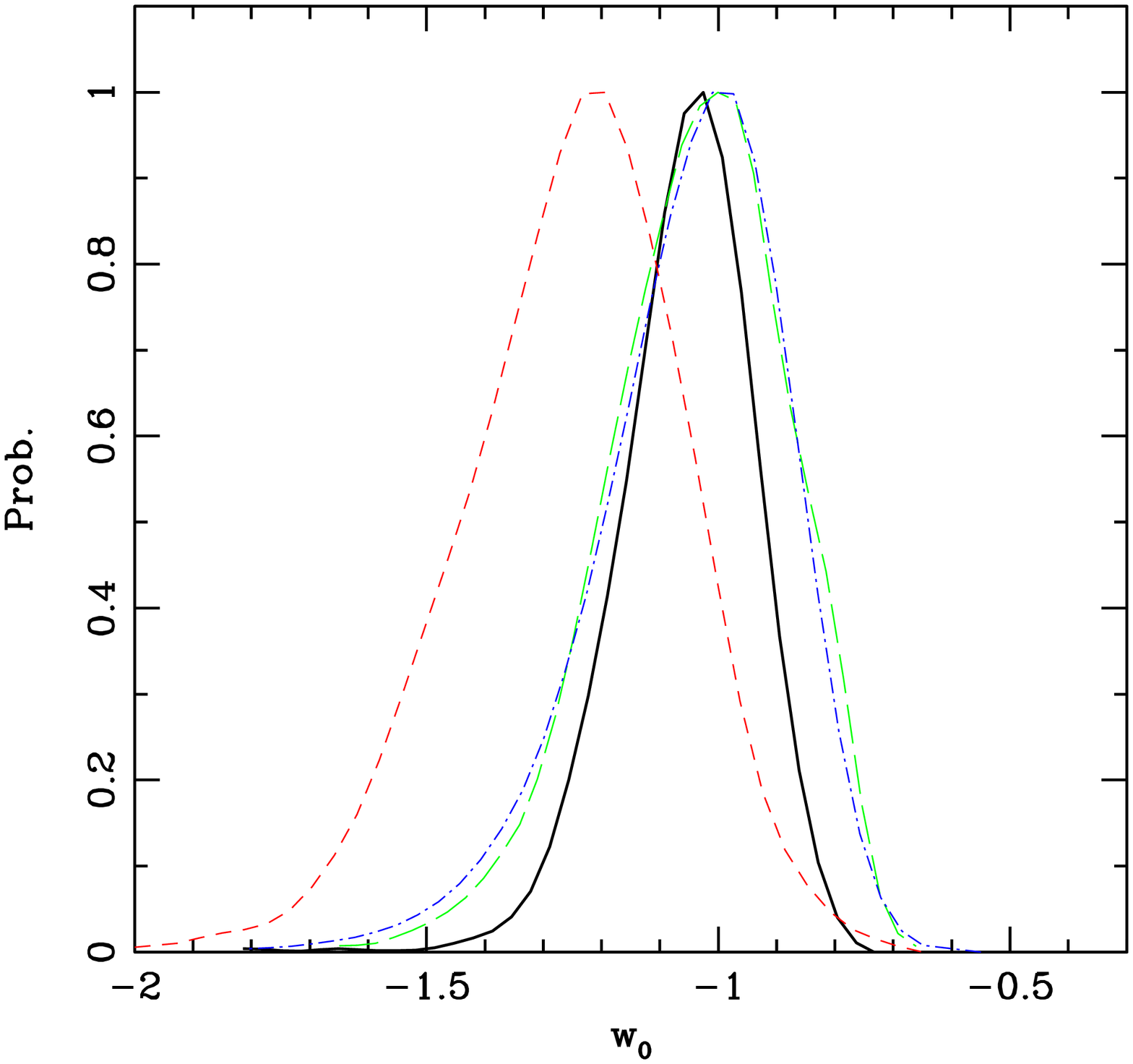}
\includegraphics[width=3.2in]{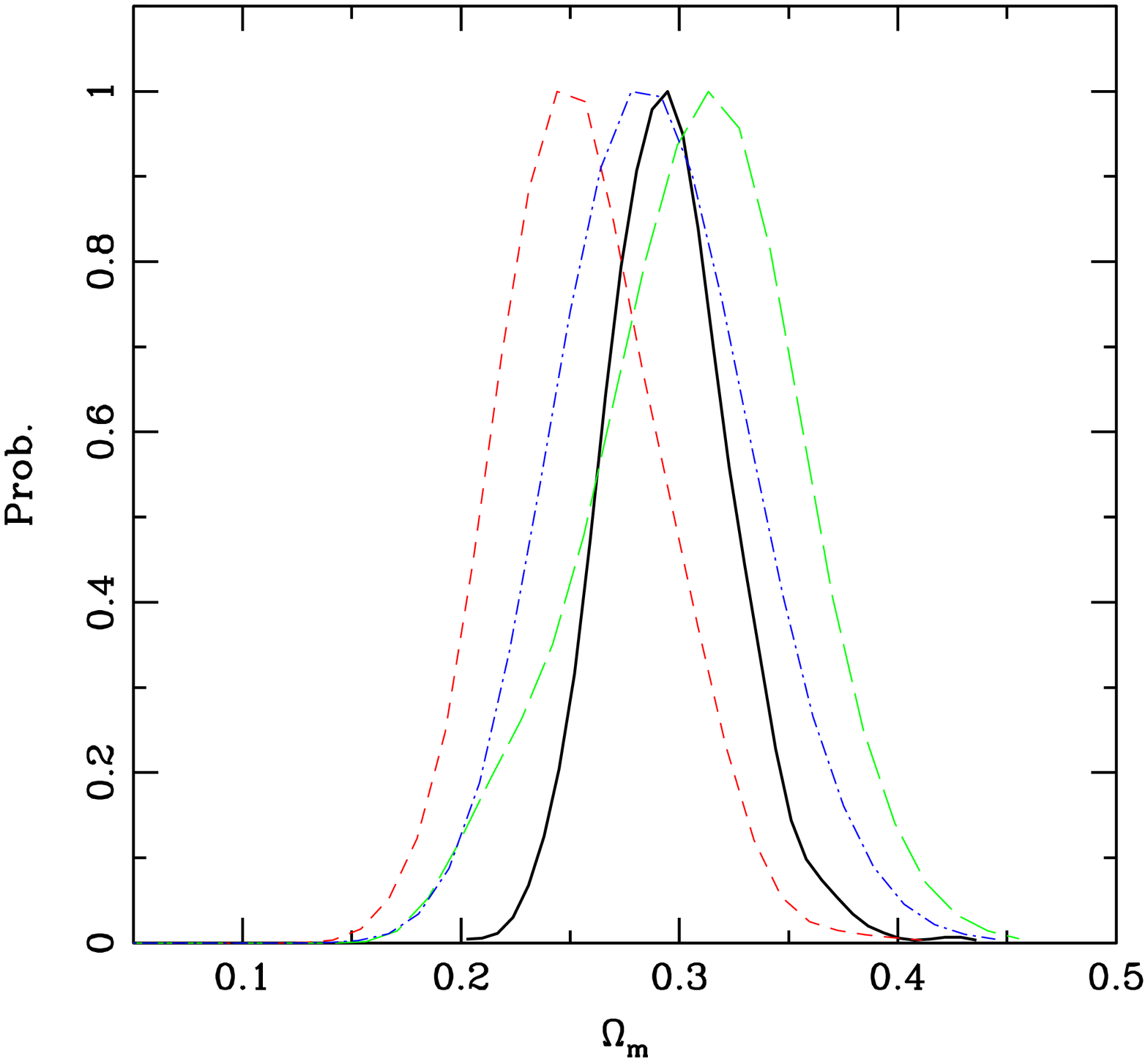}
\caption{(Left panel) The marginalised constraints on $w_{\rm 0}$
  assuming a constant dark energy equation of state. The solid line is
  for all the three data sets together, the short dashed line for
  clusters+CMB, the long-dashed line for SNIa+CMB, and the dot-dashed
  line for clusters+SNIa (with priors from HST and BBN). The right
  panel shows the marginalised constraints on $\Omega_{\rm m}$ for
  each combination of data sets.}
\label{1D}
\end{figure*}

We have included our extension of the {\sc camb} code into the
\COSMOMC~package, which provides an efficient sampling of the
posterior likelihoods using a Markov Chain Monte Carlo (MCMC)
algorithm \footnote{http://cosmologist.info/cosmomc/}~\citep{cosmomc}.
We have also included the \cite{Riess:04} supernovae sample and the
\cite{Allen:04} gas fraction data into the analysis, using the
\COSMOMC~code to calculate the posterior probability densities.

For our standard analysis we have varied nine ``cosmological''
parameters: the baryon density $\Omega_{\rm b}h^2$, the cold dark
matter density $\Omega_{\rm dm} h^2$, the Hubble constant $H_{\rm 0}$,
the reionisation redshift $z_{\rm re}$, the spectral index $n_{\rm
s}$, the amplitude of the fluctuations $A_{\rm s}$ and the dark energy
equation of state parameters $w_{\rm 0}$, $w_{\rm et}$ and $a_{\rm
t}$. We assume zero tensor components and a negligible neutrino mass.
The bias parameter $b$ associated with the X-ray cluster data is an
additional parameter in the fits. We have marginalised analytically
over the intrinsic magnitude $M$ of the supernovae. Except where
stated otherwise, our analysis assumes that the Universe is flat
($\Omega_{\rm k}=0$). For the analysis of the cluster data without the
CMB data, we use Gaussian priors on $\Omega_{\rm
b}h^2=0.0214\pm0.0020$ from Big Bang Nucleosynthesis (BBN) constraints
\citep{Kirkman:03} and $h=0.72\pm0.08$ from observations made with the
Hubble Space Telescope (HST) \citep{Freedman01}.

\begin{table}
\begin{center}
  \caption{The median values and 68.3 per cent confidence intervals
    from the analysis of the various pairs of data sets and for all
    three data sets combined, assuming a constant dark energy equation
    of state. The last column states the $\chi^{2}$ per degree of
    freedom for each combination of data sets.}
\label{tab}
\begin{tabular}{ l c c c }
\hline
Data combination    &      $w_{\rm 0}$                     &  $\Omega_{\rm m}$  & $\chi^{2}/dof$ \\
\hline

\noalign{\vskip 5pt}

Cl+CMB             &  $-1.23^{+0.17}_{-0.21}$  & $0.254^{+0.037}_{-0.034}$ & 1467.4/1378    \\

\noalign{\vskip 5pt}

Cl+SN\tiny{(+HST+BBN)}          &  $-1.04^{+0.14}_{-0.18}$  & $0.288^{+0.043}_{-0.040}$ & 210.2/178   \\
             
\noalign{\vskip 5pt}

SN+CMB                   &  $-1.04^{+0.13}_{-0.16}$  & $0.312^{+0.041}_{-0.044}$ & 1625.9/1508    \\

\noalign{\vskip 5pt}

\small{Cl+SN+CMB}          &  $ -1.05^{+0.10}_{-0.12}$ & $0.295^{+0.031}_{-0.027}$ & 1652.5/1534  \\

\hline                      
\end{tabular}
\end{center}
\end{table}

\begin{figure}
\includegraphics[width=3.2in]{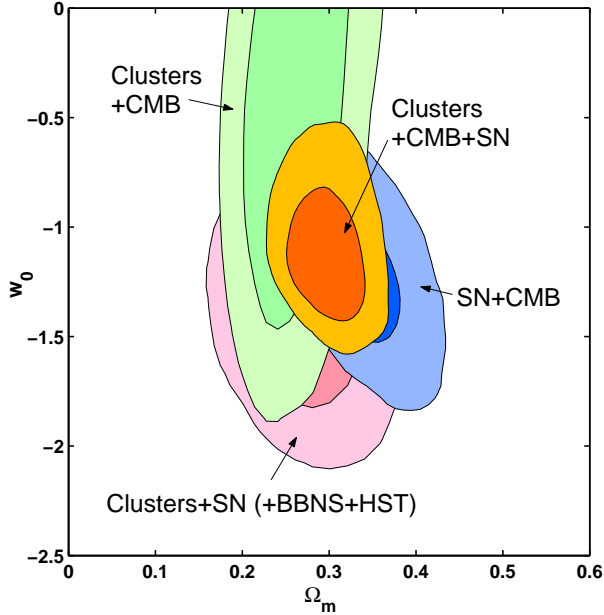}
\caption{The 68.3 and 95.4 per cent confidence limits in the
    $(\Omega_{\rm m},w_{\rm 0})$ plane for the various pairs of data
    sets and for all three data sets combined. The $z_{\rm t}=1$ dark
    energy model is assumed.}
\label{linder}
\end{figure}

For each different model+data combination, we have sampled four
independent MCMC chains. The length of these chains vary from the
simplest models of constant equation of state with around six thousand
samples per chain up to our most general model with around one hundred
and fifty thousand samples per chain. We have applied the Gelman-Rubin
criterion to test for convergence \citep{Gelman:92}.  Convergence is
assumed to be acceptable if the ratio of the between-chain and
mean-chain variances satisfies $R-1 \lt 0.1$. We have also checked
for convergence by ensuring that consistent final results are obtained
from sampling numerous small subsets of the chains.

\begin{figure*}
\includegraphics[width=2.3in]{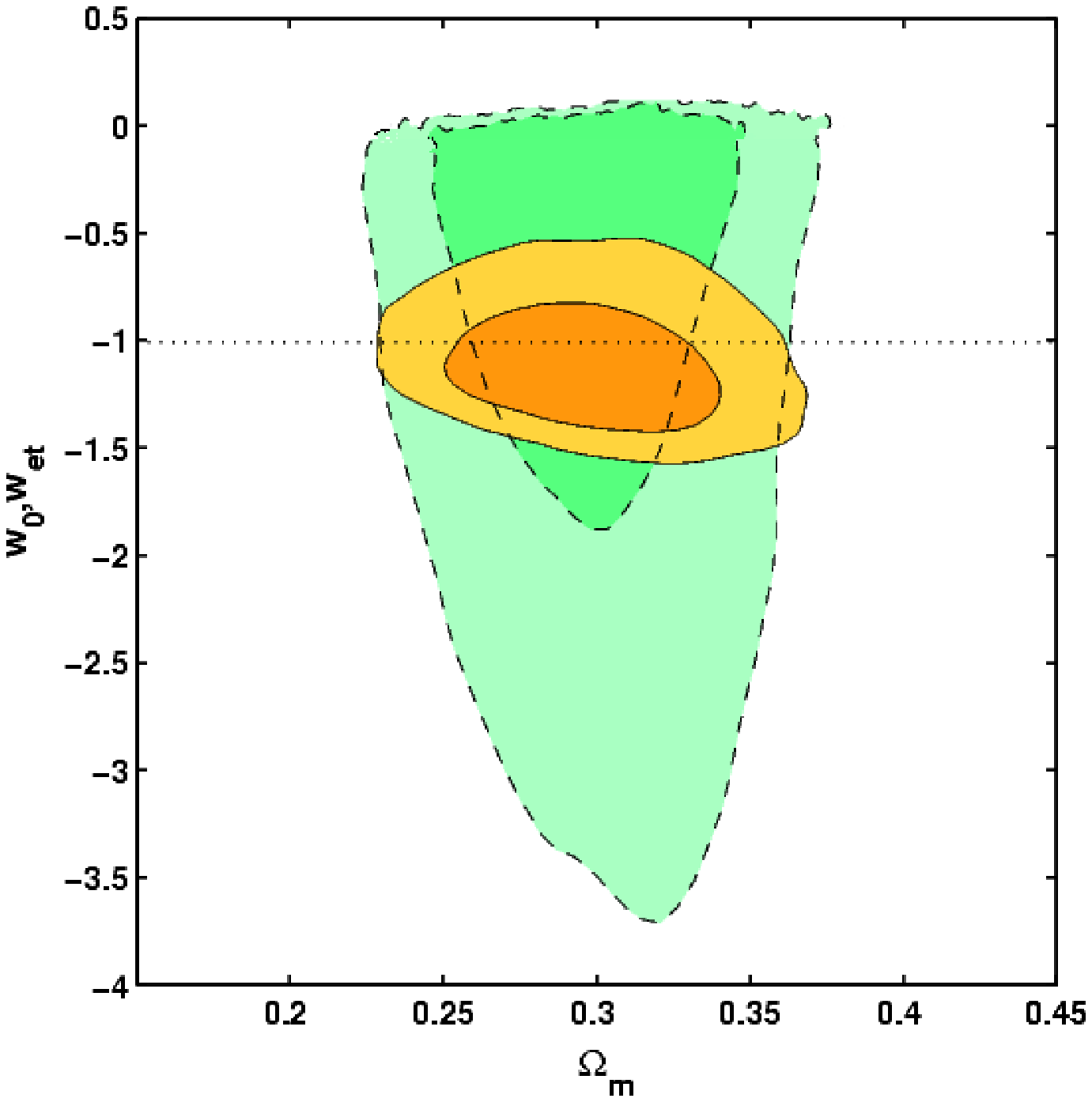}
\includegraphics[width=2.3in]{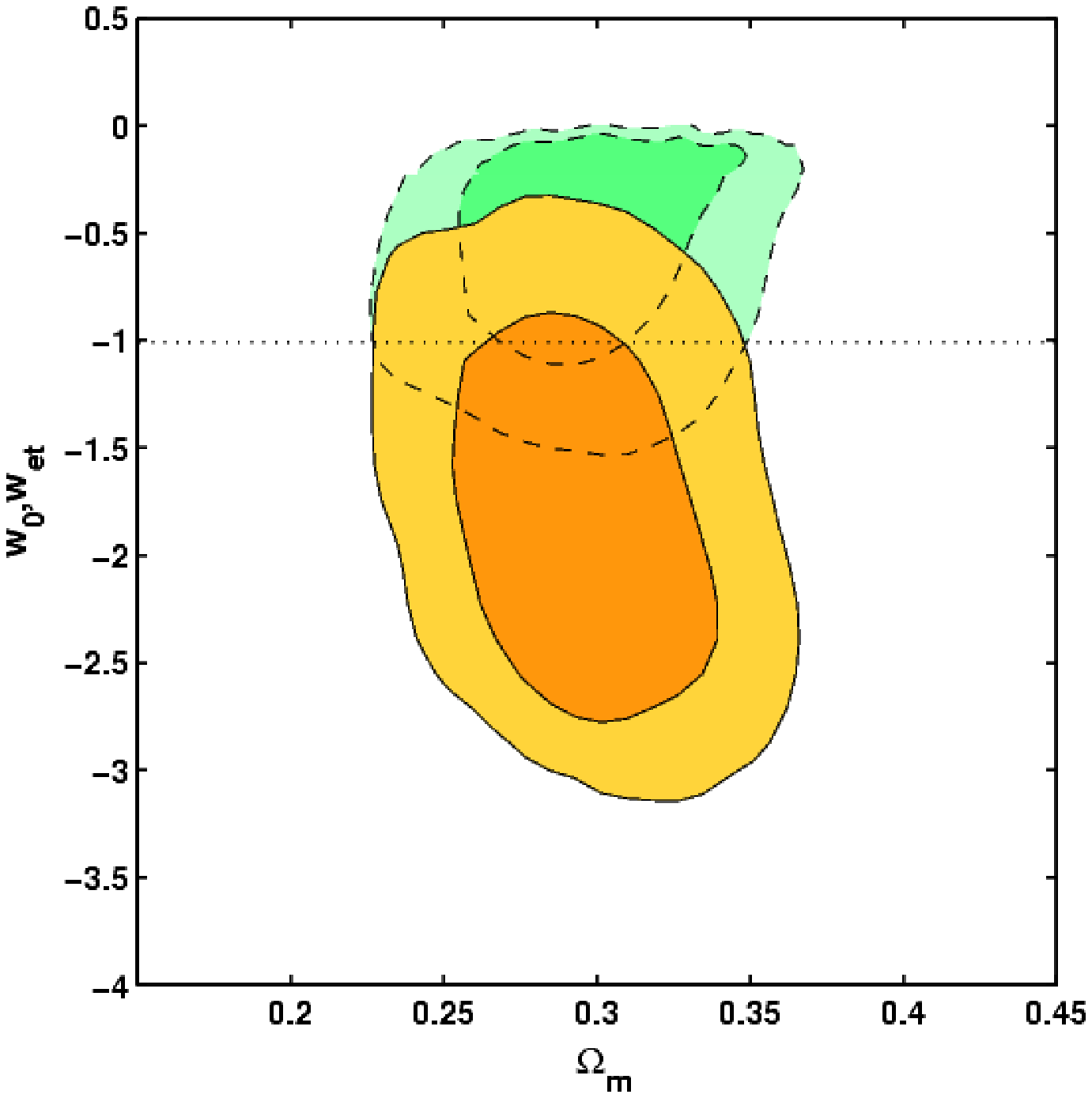}
\includegraphics[width=2.3in]{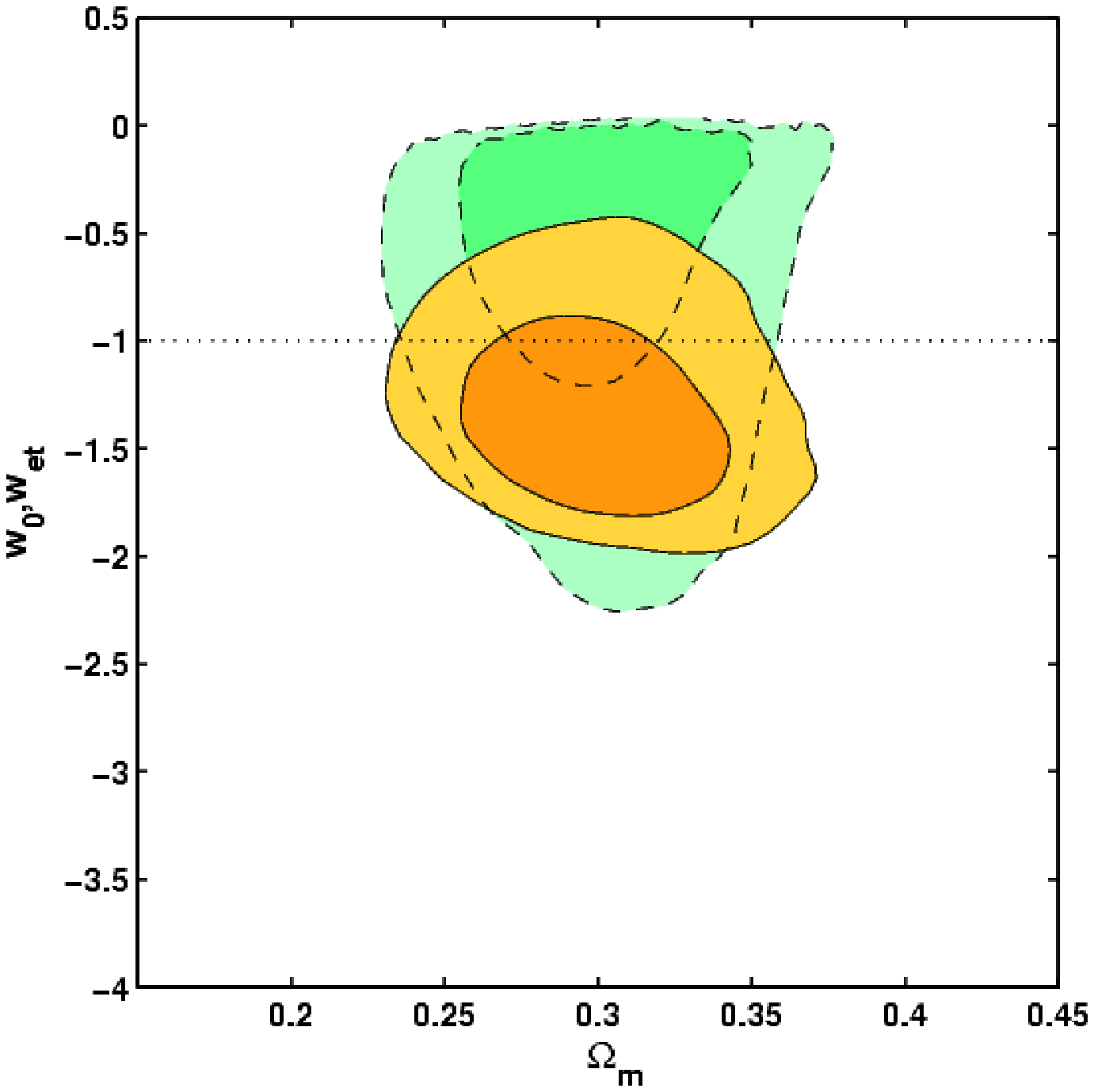}
\caption{The 68.3 and 95.4 per cent confidence limits in the
    ($\Omega_{\rm m}$;$w_{\rm 0}$,$w_{\rm et}$) plane for all three
    data sets combined using various fixed values for the transition
    redshift.  The solid lines show the results on ($\Omega_{\rm
    m}$,$w_{\rm 0}$). The dashed lines show the results on ($\Omega_{\rm
    m}$,$w_{\rm et}$). The left panel is for $z_{\rm t}=1$ ($a_{\rm
    t}=0.5$), the centre panel $z_{\rm t}=0.11$ ($a_{\rm t}=0.9$) and
    the right panel $z_{\rm t}=0.35$ ($a_{\rm t}=0.74$). The
    uncertainty in $w_{\rm et}$ is much larger than for $w_0$ in the
    left panel, which reflects the paucity of data at high
    redshifts. The $z_{\rm t}=0.35$ transition splits the cluster and
    SNIa data into similarly sized low and high redshift subsamples.
    The horizontal dotted line denotes the cosmological constant model
    ($w_{\rm 0}=w_{\rm et}=-1$).}
\label{fix}
\end{figure*}

\section{Dark Energy constraints}
\n{constraints}

We have employed a series of different parameterisations for the dark
energy equation of state: (i) $w$ constant (ii) a model with $w_{\rm
0}$ and $w_{\rm et}$ free, but with the transition redshift fixed at
$z_{\rm t}=1$ ($a_{\rm t}=0.5$) (iii) a similar model with the
transition fixed at $z_{\rm t}=0.11$ ($a_{\rm t}=0.9$) (iv) a similar
model with the transition fixed at $z_{\rm t}=0.35$ ($a_{\rm
t}=0.74$), which approximately splits the cluster and supernovae data
about their median redshifts and (v) a model in which the transition
redshift is a free parameter.

Figure~\ref{wconst} shows the constraints on $w_{\rm 0}$ and
$\Omega_{\rm m}$ for the constant dark energy equation of state model.
We see that the combination of the three data sets leads to tight
constraints on $w_{\rm 0}$ and $\Omega_{\rm m}$, which are in good
agreement with the cosmological constant scenario ($w_{\rm 0}=-1$).
This figure also demonstrates the complementary nature of the
constraints provided by the various pairs of data sets, in particular
SNIa+CMB and clusters+CMB.

The power of the combined clusters+SNIa+CMB data set is also evident
in the marginalised probability distributions for $w_{\rm 0}$ (left
panel of Figure~\ref{1D}) and $\Omega_{\rm m}$ (right panel). The
marginalised 68.3 per cent confidence limits on $w_{\rm 0}$ and
$\Omega_{\rm m}$ for the various data combinations, assuming a
constant dark energy equation of state, are summarised in Table
\ref{tab}.

Figure~\ref{linder} shows the results obtained using the $z_{\rm t}=1$
dark energy model.  Comparison of Figs~\ref{wconst} and~\ref{linder}
shows how the joint confidence contours on ($\Omega_{\rm m}$,$w_{\rm 0}$)
open up when $w_{\rm et}$ is introduced as an additional free
parameter. This is particularly prominent for the clusters+CMB
combination where the $w_{\rm 0}$ region is extended into the positive
branch due to the degeneracy between $w_{\rm 0}$ and $w_{\rm et}$,
discussed below.  The results on $w_{\rm 0}$ and $w_{\rm et}$ for the
$z_{\rm t}=1$ model are shown in the left panel of Figure~\ref{fix}.
The figure shows how, for this model, the present data constrain
$w_{\rm 0}$ more tightly than $w_{\rm et}$. This simply reflects the
paucity of cluster and SNIa data at redshifts beyond $z_{\rm t}=1$.
Note that the CMB data provide an upper limit of $w_{\rm et} \lsim 0$
at high redshifts; for $w_{\rm et} \gt 0$, the dark energy component
would become significant at early times, causing modifications to the
predicted CMB anisotropy spectrum.

\begin{figure}
\includegraphics[width=3.2in]{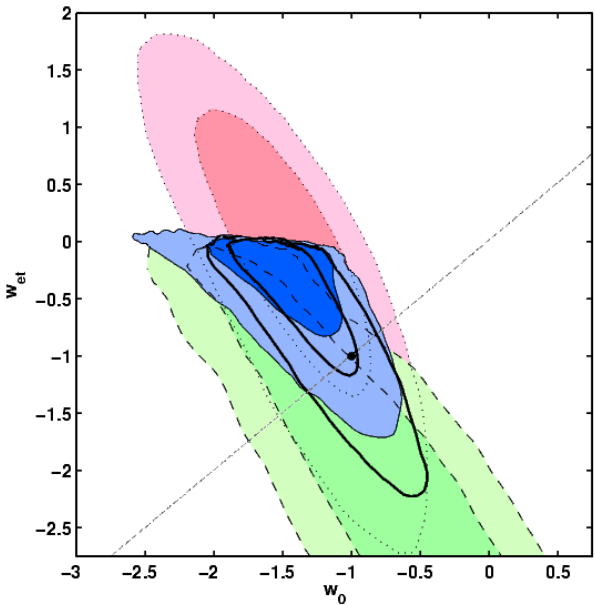}
\caption{The 68.3 and 95.4 per cent confidence limits in the ($w_{\rm
    0}$,$w_{\rm et}$) plane obtained using the SNIa+CMB (blue, thin
    solid), Clusters+CMB (green, long--dashed) and
    Clusters+SNIa(+BBN+HST) (magenta, dotted) data sets. The bold
    lines show the results for all three data sets combined and the
    dark circle marks the cosmological constant model. The dashed line
    shows the no evolution models ($w_{\rm 0}=w_{\rm et}$). The
    transition redshift is fixed to $z_{\rm t}=0.35$ in all cases.}
\label{split2}
\end{figure}

\begin{table*}
\begin{center}
  \caption{The median parameter values and values at the peaks of the
    marginalised probability distributions (and 68.3 per cent
    confidence intervals) for the various dark energy
    parameterisations, using all three data sets combined. Results are
    listed for both flat and non-flat priors. The last column states
    the $\chi^{2}$ per degree of freedom for each parameterisation. }
\label{tab2}
\begin{tabular}{ l l l l l l l c }
\hline
equation of state          &   median $w_{\rm 0}$       &  median $w_{\rm et}$           &  median $\Omega_{\rm m}$         & Peak $w_{\rm 0}$    & Peak $w_{\rm et}$ & Peak $\Omega_{\rm m}$ & $\chi^{2}/dof$ \\
\hline

\noalign{\vskip 5pt}

constant (flat) &$ -1.05^{+0.10}_{-0.12}$&  -        & $0.295^{+0.031}_{-0.027}$& $-1.04\pm0.10$ &-      &$0.293\pm0.028$ & 1652.5/1534 \\

\noalign{\vskip 5pt}

constant (non flat)&$-1.09^{+0.12}_{-0.15}$&  -      & $0.314^{+0.040}_{-0.036}$& $-1.02^{+0.08}_{-0.20}$ &-      &$0.31^{+0.04}_{-0.05}$ & 1651.3/1533 \\

\noalign{\vskip 5pt}

$z_{\rm t} =1$ (flat)      &$-1.10^{+0.23}_{-0.19}$&$-0.87^{+0.61}_{-1.10}$& $0.300^{+0.029}_{-0.028}$&$-1.16^{+0.22}_{-0.19}$&$-0.05^{+0.09}_{-1.17}$&$0.30\pm0.03$ &  1650.6/1533 \\

\noalign{\vskip 5pt}

$z_{\rm t} = 1$ (non flat)  &$-1.08^{+0.30}_{-0.23}$         & $-1.23^{+0.86}_{-2.06}$         & $0.328^{+0.046}_{-0.040}$         &$-1.14^{+0.31}_{-0.21}$&$-0.09^{+0.12}_{-2.16}$&$0.32^{+0.04}_{-0.05}$ & 1648.1/1532 \\

\noalign{\vskip 5pt}

split, $z_{\rm t} = 0.35$ (flat)       &$-1.30^{+0.34}_{-0.28}$&$-0.61^{+0.40}_{-0.62}$& $0.300^{+0.028}_{-0.027}$         &$-1.49^{+0.41}_{-0.20}$&$-0.10^{+0.09}_{-0.71}$&$0.30\pm0.03$  & 1649.0/1533\\

\noalign{\vskip 5pt}
             
arbitrary $z_{\rm t}$ (flat) &$-1.27^{+0.33}_{-0.39}$         & $-0.66^{+0.43}_{-0.62}$         & $0.299^{+0.029}_{-0.027}$         &$-1.23^{+0.34}_{-0.46}$&$-0.12^{+0.11}_{-0.76}$&$0.298\pm0.028$ & 1648.9/1532\\

\noalign{\vskip 5pt}

\hline

\end{tabular}
\end{center}
\end{table*}

It is important to recognise that the choice of transition redshift,
$z_{\rm t} = 1$, described above is arbitrary. We have therefore
examined the constraints obtained for other values of $z_{\rm t}$
($a_{\rm t}$). The centre panel of Figure~\ref{fix} shows the results
using a late transition model with $z_{\rm t}=0.11$ ($a_{\rm t}=0.9$).
We see that the cosmological constant ($w_{\rm 0}=w_{\rm et}=-1$)
again lies within the allowed 68.3 per cent confidence ($1\sigma$)
regions. Unsurprisingly, the constraints on $w_{\rm et}$ in the late
transition case are better than for the $z_{\rm t}=1$ model,
reflecting the presence of more cluster and SNIa data beyond the
transition redshift. Naturally, this at the expense of a weaker
constraint on $w_{\rm 0}$.

If we select a transition redshift close to the median redshift for
the SNIa and cluster data sets, one might expect to obtain comparable
constraints on $w_{\rm 0}$ and $w_{\rm et}$. In principle, this
approach could provide improved sensitivity when searching for
evolution in the equation of state parameter. (In detail, we expect
the constraints on $w_{\rm 0}$ to be slightly better than those for
$w_{\rm et}$ using the median redshift model, since the precision of
the individual cluster and supernova measurements are lower at high
redshifts.)  The right panel of Figure~\ref{fix} shows the results
obtained fixing $z_{\rm t} = 0.35$ ($a_{\rm t}=0.74$), a redshift
close to the median redshift for both the cluster and SNIa data sets.
In this case the uncertainties on $w_{\rm 0}$ and $w_{\rm et}$ are
indeed similar and the combined size of the confidence regions is
reduced. However, the cosmological constant remains an acceptable
description of the data. The marginalised results on $w_{\rm 0}$,
$w_{\rm et}$ and $\Omega_{\rm m}$ are summarised in Table~\ref{tab2}.

Within the context of searching for evolution in the dark energy
equation of state, it is interesting to note the constraints that
arise from the combinations of SNIa+CMB, clusters+CMB and
clusters+SNIa(+BBN+HST) data separately. Figure~\ref{split2} shows the
results in the ($w_{\rm 0},w_{\rm et}$) plane for the three pairs of
data sets and for all three data sets combined, using the $z_{\rm t} =
0.35$ dark energy model. We see that the combination of SNIa+CMB data
provides marginal evidence for evolution in the equation of state, in
that the 68.3 per cent confidence contours exclude the no evolution
line ($w_{\rm 0}=w_{\rm et}$). However, the clusters+CMB and
clusters+SNIa data are consistent with the cosmological constant at
the 68.3 per cent level, and the effect of combining all three data
sets (bold contours in Figure~\ref{split2}) is to remove the marginal
evidence for evolution hinted at in the SNIa+CMB data alone.

\begin{figure}
\includegraphics[width=3.2in,height=3.2in]{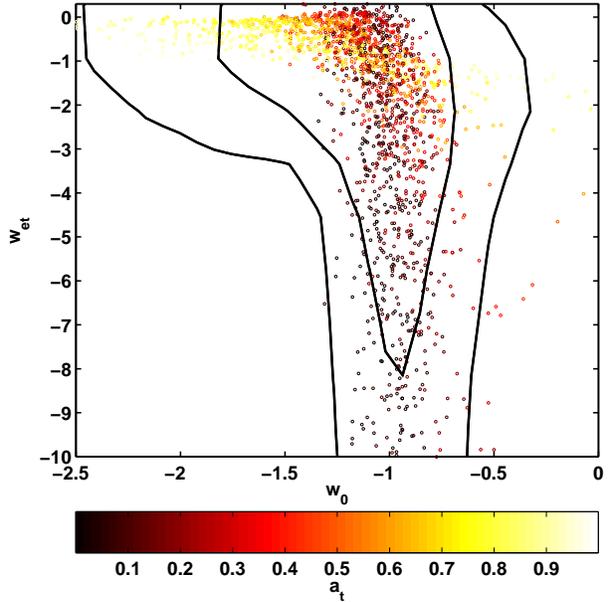}
\caption{The distribution of MCMC samples in the ($w_{\rm 0}$,$w_{\rm
  et}$) plane for the case of a transition scale factor allowed to vary
  over the range $0\lt a_{\rm t}\lt 1$.  The colours indicate the
  value of $a_{\rm t}$.}
\label{fig:at2d}
\end{figure} 

\begin{figure*}
\includegraphics[width=3.2in]{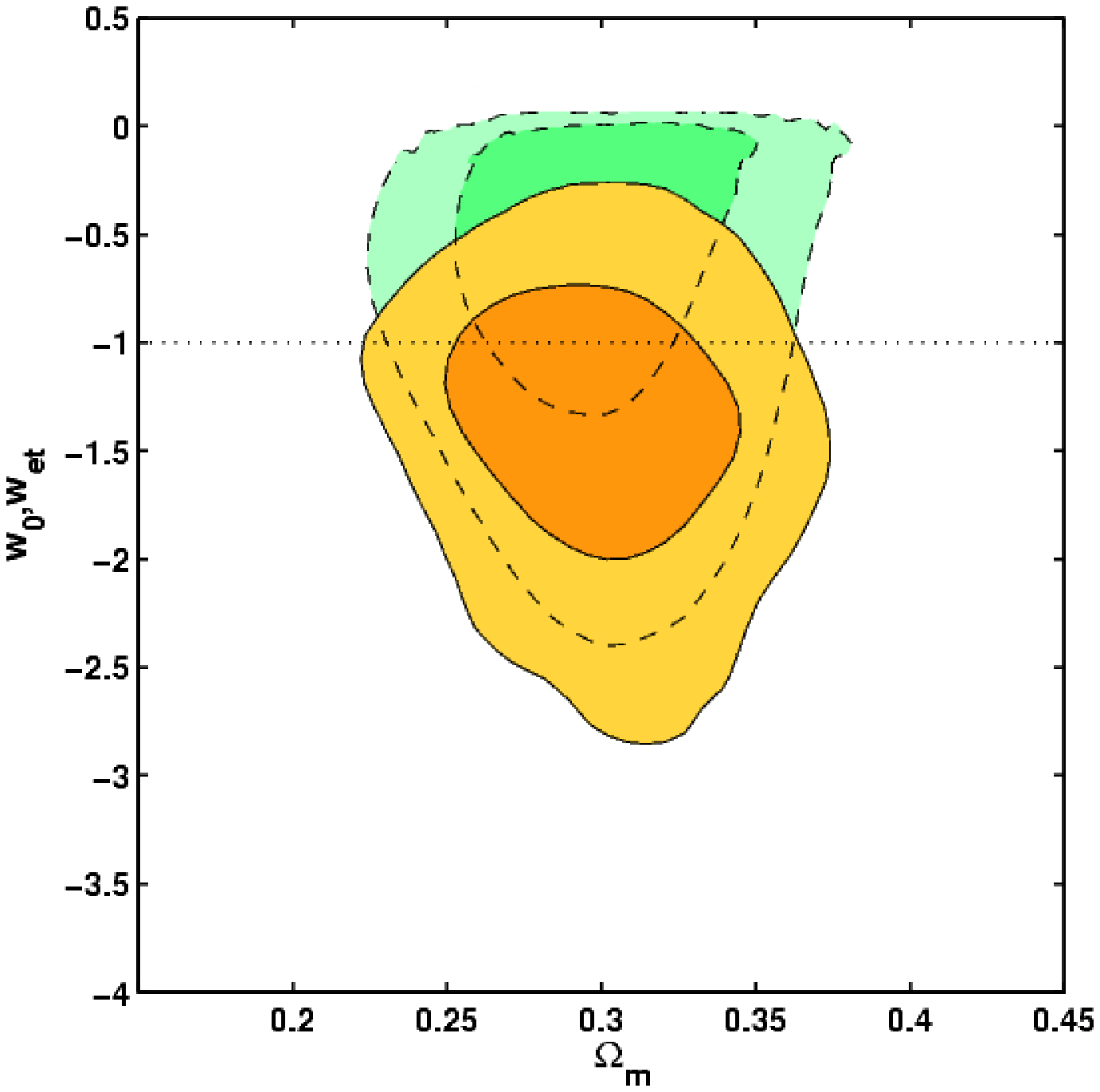}
\includegraphics[width=3.2in]{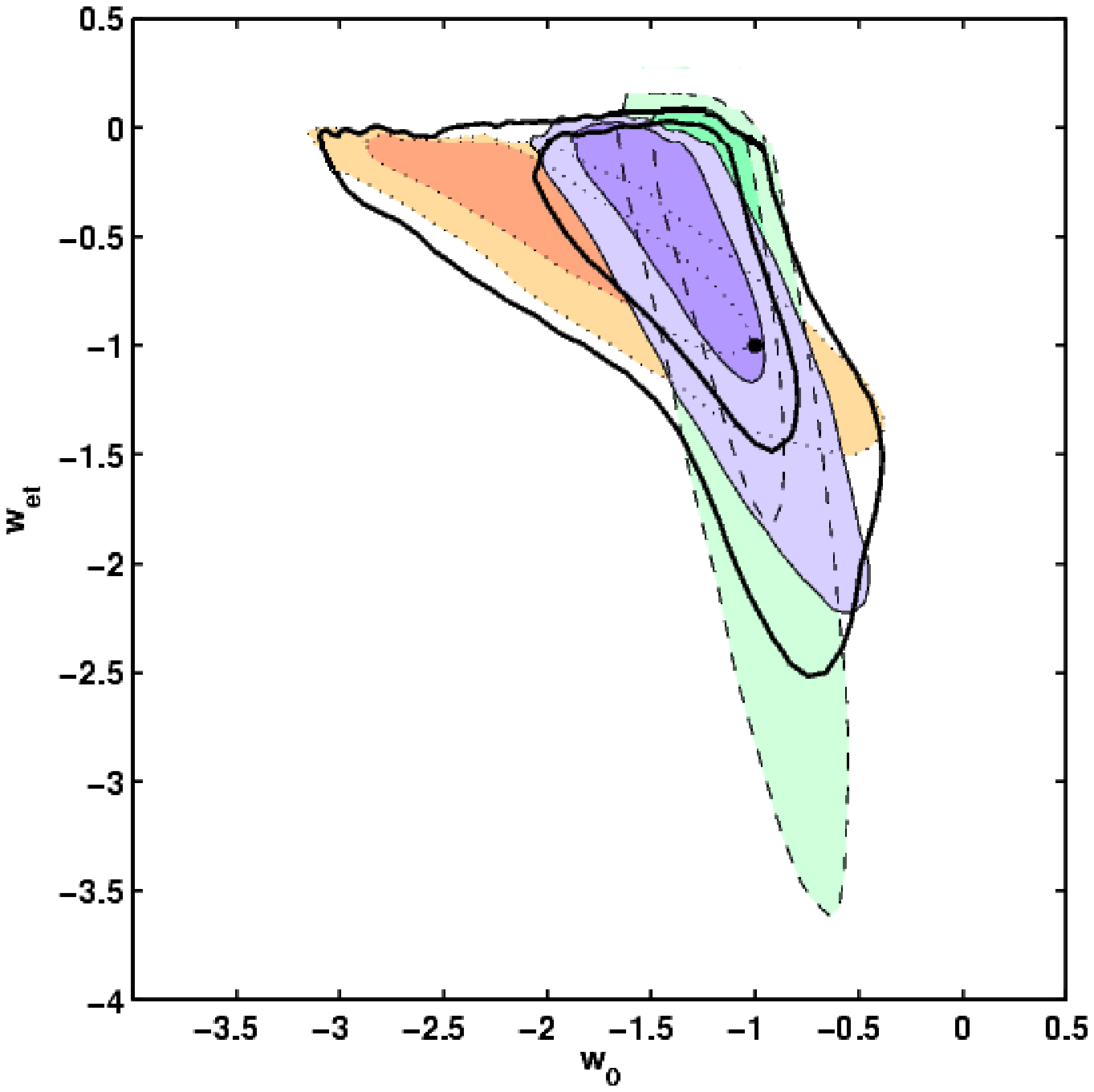}
\caption{(Left panel) The 68.3 and 95.4 per cent confidence limits in
the ($\Omega_{\rm m}$;$w_{\rm 0}$,$w_{\rm et}$) plane for a general
dark energy model with the transition scale factor allowed to vary
over the range $0.5\lt a_{\rm t}\lt 0.95$. The solid lines show the
results on ($\Omega_{\rm m}$,$w_{\rm 0}$). The dashed lines show the
results on ($\Omega_{\rm m}$,$w_{\rm et}$). The horizontal dotted line
denotes the cosmological constant model ($w_{\rm 0}=w_{\rm
et}=-1$). (Right panel) The constraints in the ($w_{\rm 0}$,$w_{\rm
et}$) plane for the above model (bold, solid lines). For comparison,
the constraints obtained for the various fixed transition redshifts
are also shown: $z_{\rm t} = 1.0$ (green, dashed), $z_{\rm t} = 0.35$
(blue, thin solid), $z_{\rm t} = 0.11$ (red, dotted). The dark circle
marks the cosmological constant model.}
\label{fig:atpriors}
\end{figure*} 

Figure~\ref{split2} clearly shows the degeneracies between $w_{\rm 0}$
and $w_{\rm et}$ for the different data combinations, and demonstrates
the importance of including the CMB data when attempting to obtain the
best constraints on $w_{\rm 0}$ and $w_{\rm et}$. Comparing the dotted
contours [clusters+SNIa+(BBN+HST)] with the bold contours
[clusters+SNIa+CMB] one sees that as well as providing a tight upper
limit on $w_{\rm et}$ as discussed above, the inclusion of the CMB
data also leads to tighter constraints on $w_{\rm 0}$ and the
exclusion of large negative values for $w_{\rm et}$.

The most general dark energy model we have examined includes $w_{\rm
  0}$, $w_{\rm et}$ and the transition scale factor, $a_{\rm t}$, as
free parameters. In the first case, $a_{\rm t}$ was allowed to take
any value in the range $0\lt a_{\rm t}\lt 1$. The distribution of MCMC
samples from this analysis is shown in Figure \ref{fig:at2d}. This
figure re-emphasises the point that when the transition in the dark
energy equation of state occurs at late times (low redshifts; light
coloured sample points) $w_{\rm et}$ is confined to a relatively
narrow band ($-2 \lsim w_{\rm et} \lsim 0$) and the constraint on
$w_{\rm 0}$ is poor ($-2.5 \lsim w_{\rm 0} \lsim -0.5$). When the
transition occurs earlier (at high redshifts; darker points), the
constraint on $w_{\rm 0}$ is improved and the constraint on $w_{\rm
  et}$ is weakened. In this case, some sample points even populate the
region beyond $w_{\rm et}\gt0$. [If the transition occurs at
sufficiently early times, even the CMB data cannot provide a tight
limit on $w_{\rm et}$.  It is likely, however, that an early equation
of state with $w \gsim 1/3$ violates BBN bounds, if the dark energy
component is significant at the time when the BBN species freeze out
\citep{Bean:01}.]

\begin{figure}
\includegraphics[width=3.2in]{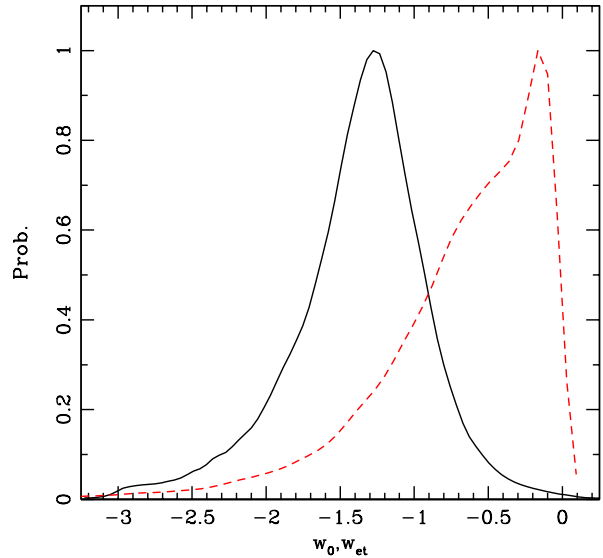}
\caption{The marginalised results on $w_{\rm 0}$ (solid line) and
  $w_{\rm et}$ (dashed line) for the general dark energy model with
  the transition scale factor allowed to vary over the range $0.5\lt
  a_{\rm t}\lt 0.95$ (c.f. Fig~\ref{fig:atpriors}).}
\label{fig:at1d}
\end{figure}

It is clear from Fig~\ref{fig:at2d} that interesting constraints on
both $w_{\rm 0}$ and $w_{\rm et}$ can only be obtained when the
transition redshift is restricted to lie within the range spanned by
the cluster and SNIa data. Otherwise large peaks in the marginalised
probability distributions will occur that will simply reflect an
inability to distinguish between models with transition redshifts
beyond the range of the present data.  For this reason, we have
carried out a second analysis in which $a_{\rm t}$ was allowed to vary
only over the range $0.5\lt a_{\rm t}\lt 0.95$; a sensible compromise
given the current cluster and SNIa data.  Fig~\ref{fig:atpriors} shows
the confidence contours in the ($\Omega_{\rm m}$;$w_{\rm 0}$,$w_{\rm
et}$) (left panel) and ($w_{\rm 0}$,$w_{\rm et}$) planes (right panel,
bold-solid lines). The marginalised results on $w_{\rm 0}$ and
$w_{\rm et}$ are shown in Fig~\ref{fig:at1d}. Again, the results
obtained with our most general dark energy model are consistent with a
cosmological constant.

As mentioned above, our analysis accounts for the effects of spatial
fluctuations in the dark energy component. Most previous studies have
not accounted for these perturbations, despite the fact that this
leads to violation of energy--momentum conservation whenever dark
energy is not a cosmological constant \citep{wayne,Caldwell:05}.  It
is important to ask whether the inclusion of these perturbations has a
significant effect on the results; it has been argued by some authors
that dark energy perturbations can be neglected if the equation of
state remains around the cosmological constant value. However, we find
that for an evolving equation of state, neglecting the effects of such
perturbations can lead to spuriously tight constraints on the dark
energy parameters.

For a constant equation of state, \cite{Weller:03} showed that the
inclusion of dark energy perturbations leads to an opening up of the
($\Omega_{\rm m}$,$w_{\rm 0}$) contours, allowing more negative
values of $w_{\rm 0}$. Repeating their analysis using our three data
sets (clusters+SNIa+CMB) we measure a reduced effect, due to the
complementary nature of our data sets. Neglecting the effects of dark
energy perturbations leads to only a small shift in the marginalized
probability distribution for $w_{\rm 0}$ and slightly tighter
constraints ($w_{\rm 0}=-0.99^{+0.09}_{-0.11}$; see Table~\ref{tab}
for the results obtained including perturbations).

\begin{figure}
\includegraphics[width=3.2in,height=3.2in]{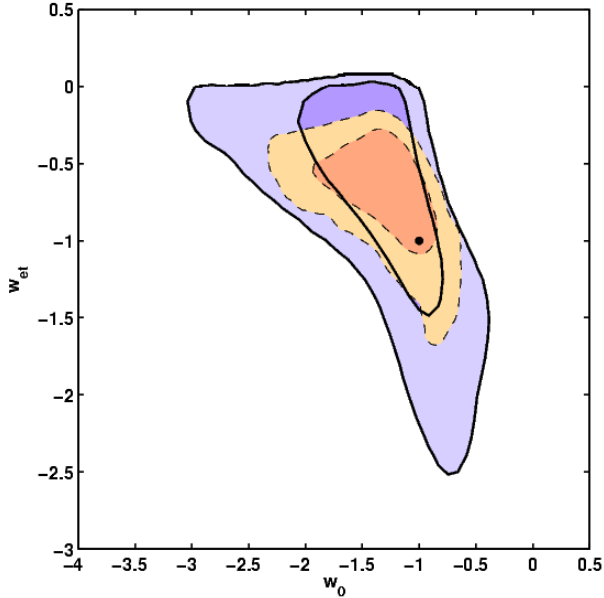}
\caption{The 68.3 and 95.4 per cent confidence limits in the ($w_{\rm
0}$,$w_{\rm et}$) plane obtained from analyses which account for
(blue, solid contours; as in Fig~\ref{fig:atpriors}) or incorrectly
neglect (red, dashed contours) the effects of dark energy
perturbations. The model used incorporates a free transition redshift
for the dark energy equation of state and is fitted to all three data
sets: clusters+SNIa+CMB.}
\label{evol:nopert}
\end{figure} 

\begin{figure}
\includegraphics[width=3.2in,height=3.2in]{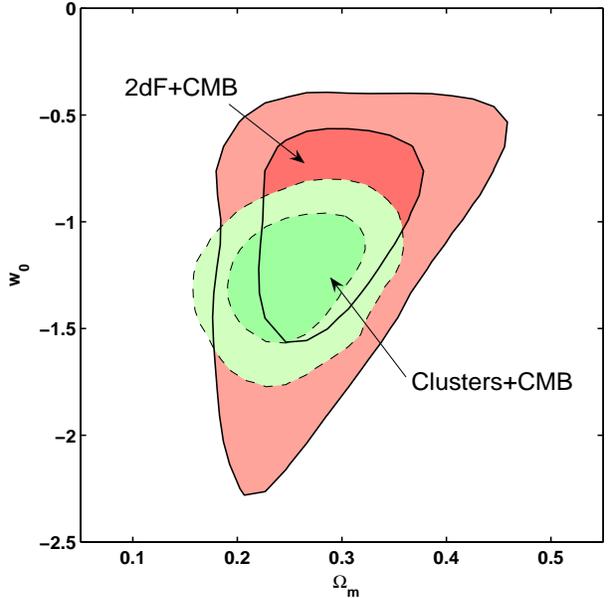}
\caption{The 68.3 and 95.4 per cent confidence limits in the
    ($\Omega_{\rm m}$,$w_{\rm 0}$) plane assuming a constant dark
    energy equation of state and combining CMB data either with the
    2dF galaxy redshift survey \citep{Percival02} (red, solid
    contours) or with the X-ray cluster gas mass fraction data (green,
    dashed contours).}
\label{2dfwconst}
\end{figure} 

For our most general, evolving dark energy model, however, the effects
of perturbations in the dark energy component are more important and
neglecting them can lead to spuriously tight constraints.
Figure~\ref{evol:nopert} compares the results in the ($w_{\rm
0}$,$w_{\rm et}$) plane obtained when including dark energy
perturbations (solid contours) or neglecting them (dashed
contours). When the effects of dark energy perturbations are wrongly
ignored, we obtain spuriously tight constraints on $w_{\rm
0}=-1.25^{+0.25}_{-0.34}$ and especially $w_{\rm
et}=-0.67^{+0.21}_{-0.30}$; the apparent uncertainties in the
latter are reduced by a factor of $\sim 2$ from the values in
Table~\ref{tab2}. Figure~\ref{evol:nopert} also shows that when we
incorrectly neglect the effects of dark energy perturbations, the
sharp boundary provided by the CMB data set around $w_{\rm et}\sim 0$
is not reached.

Finally it is interesting to compare the constraints on the matter
content and dark energy parameters from this study with other work
where the CMB and SNIa data have been combined with independent
constraints from galaxy redshift surveys (e.g. \cite{Spergel03},
\cite{Tegmark:sdss}). The cluster X-ray gas mass fraction and galaxy
redshift survey data break parameter degeneracies in the CMB data in
similar ways. This is shown in Figure~\ref{2dfwconst} which compares
the results for a constant equation of state model using the
Cluster+CMB data (green, dashed contours) or when combining the CMB
data with constraints from the 2dF galaxy redshift survey
(\cite{Percival02}; red, solid contours. The are the same constraints
used by \cite{Spergel03}.)  It is clear from the figure that the
inclusion of the cluster gas fraction data leads to tighter
constraints on $\Omega_{\rm m}$ and, especially, $w_{\rm 0}$. However,
when the SNIa data are also included, the constraints from the
clusters+CMB+SNIa and 2dF+CMB+SNIa data sets are similar, due again to
the complementary nature of the various data sets. For the
2dF+CMB+SNIa data, we obtain $w_{\rm 0}= -1.02^{+0.12}_{-0.13}$ and
$\Omega_{\rm m}= 0.304\pm 0.031$, which are comparable to the values
in Table~\ref{tab}. Note that the 2dF+CMB constraints shown in
Figure~\ref{2dfwconst} are somewhat weaker than those presented by
\cite{Spergel03}. This is due primarily to the inclusion of dark
energy perturbations in the present study. \cite{Spergel03} also
employ a prior $\tau<0.3$ which has a small effect on the
contours. Such a prior is not included here.

\begin{figure*}
\includegraphics[width=3.2in]{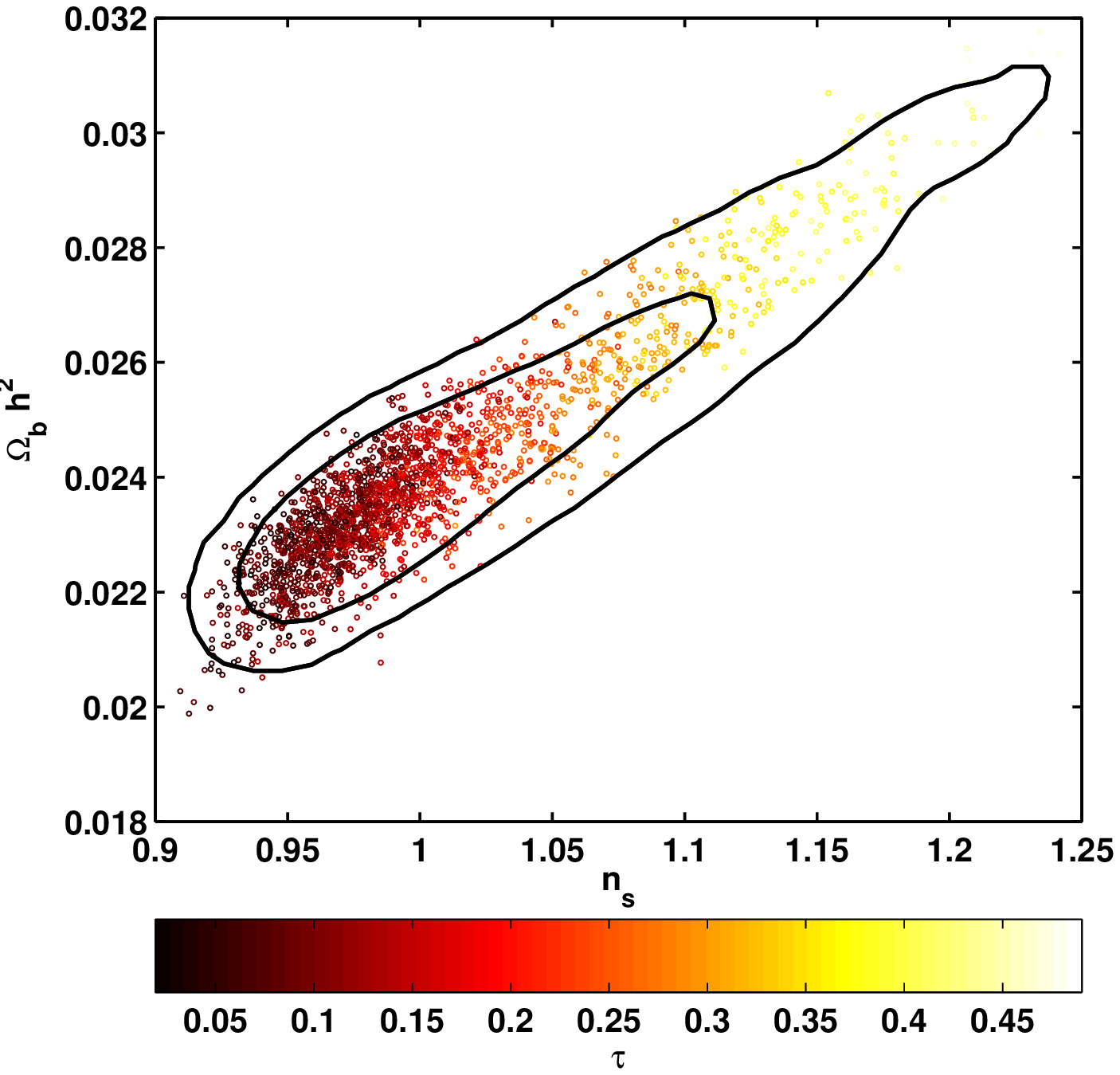}
\includegraphics[width=3.2in]{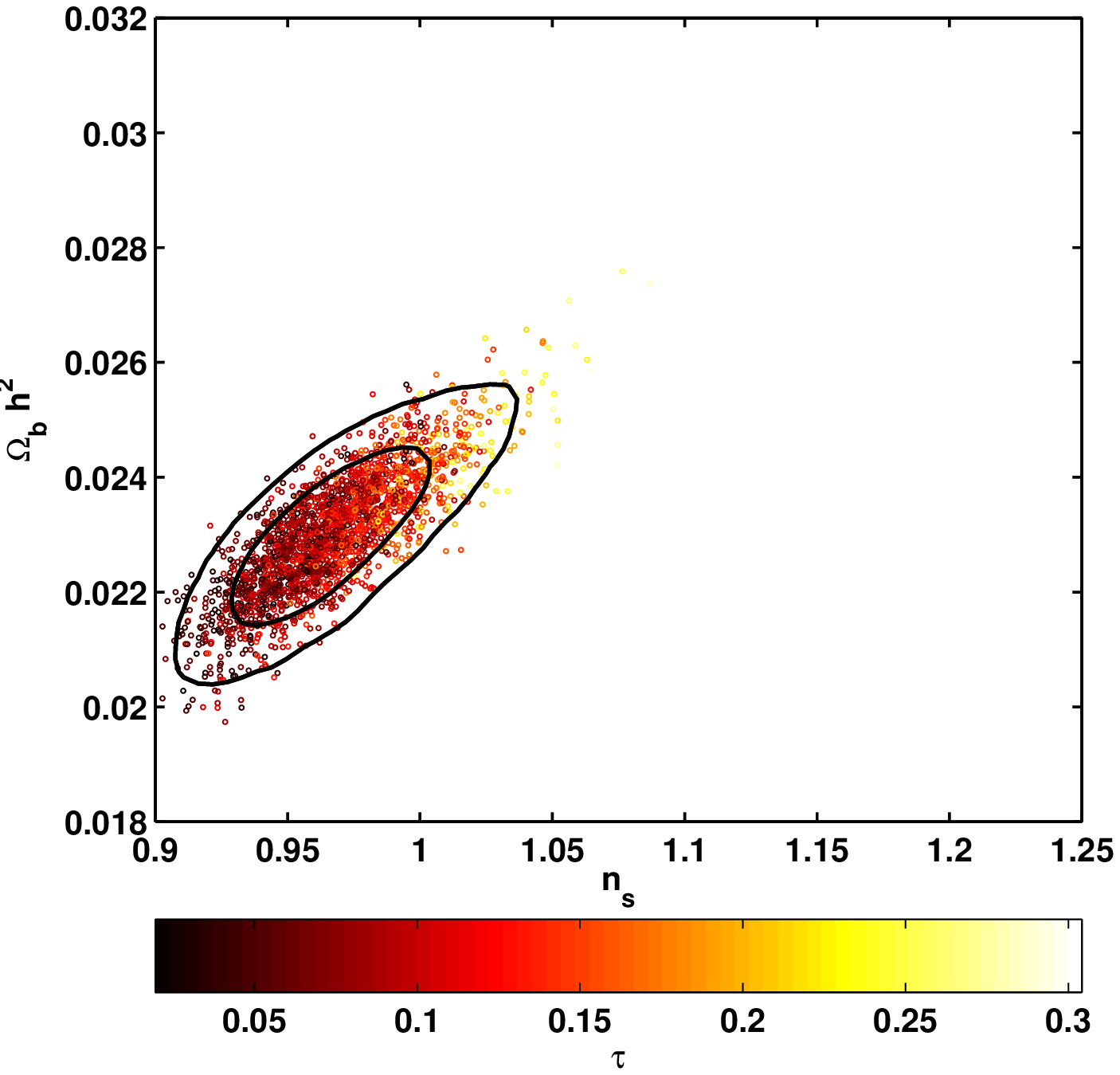}
\caption{The 68.3 and 95.4 per cent confidence limits in the
  $(n_{s},\Omega_{\rm b}h^2)$ plane from the analyses of SNIa+CMB data
  (left panel) and clusters+CMB data (right panel) using the dark
  energy model with $z_{\rm t}=0.35$. Also shown are the thinned
  samples, where the colours correspond to the value of $\tau$. Note
  how the combination of clusters+CMB data alleviates the degeneracies
  between these parameters.}
\label{degen}
\end{figure*}

\begin{figure*}
\includegraphics[width=3.2in]{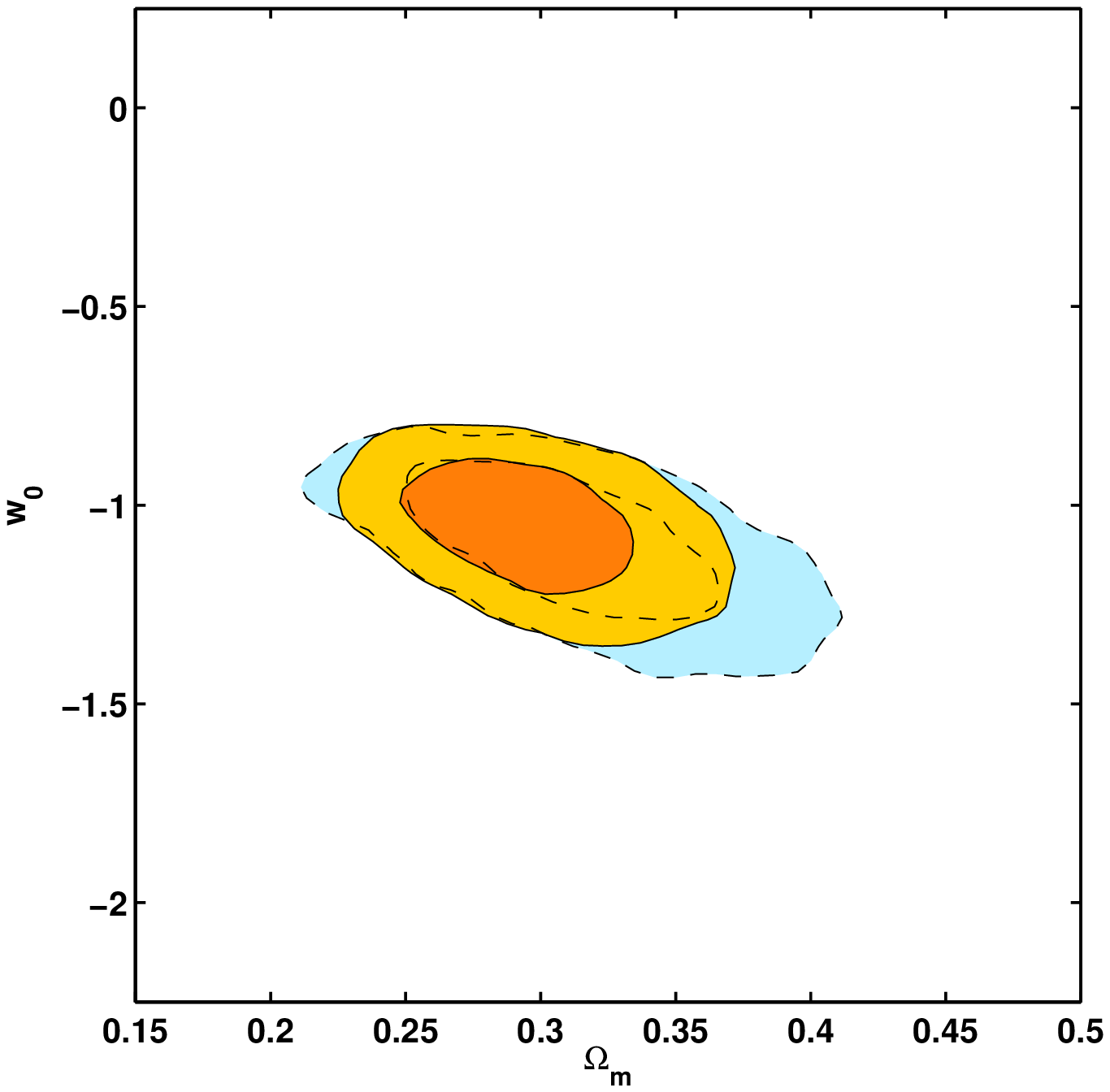}
\includegraphics[width=3.2in]{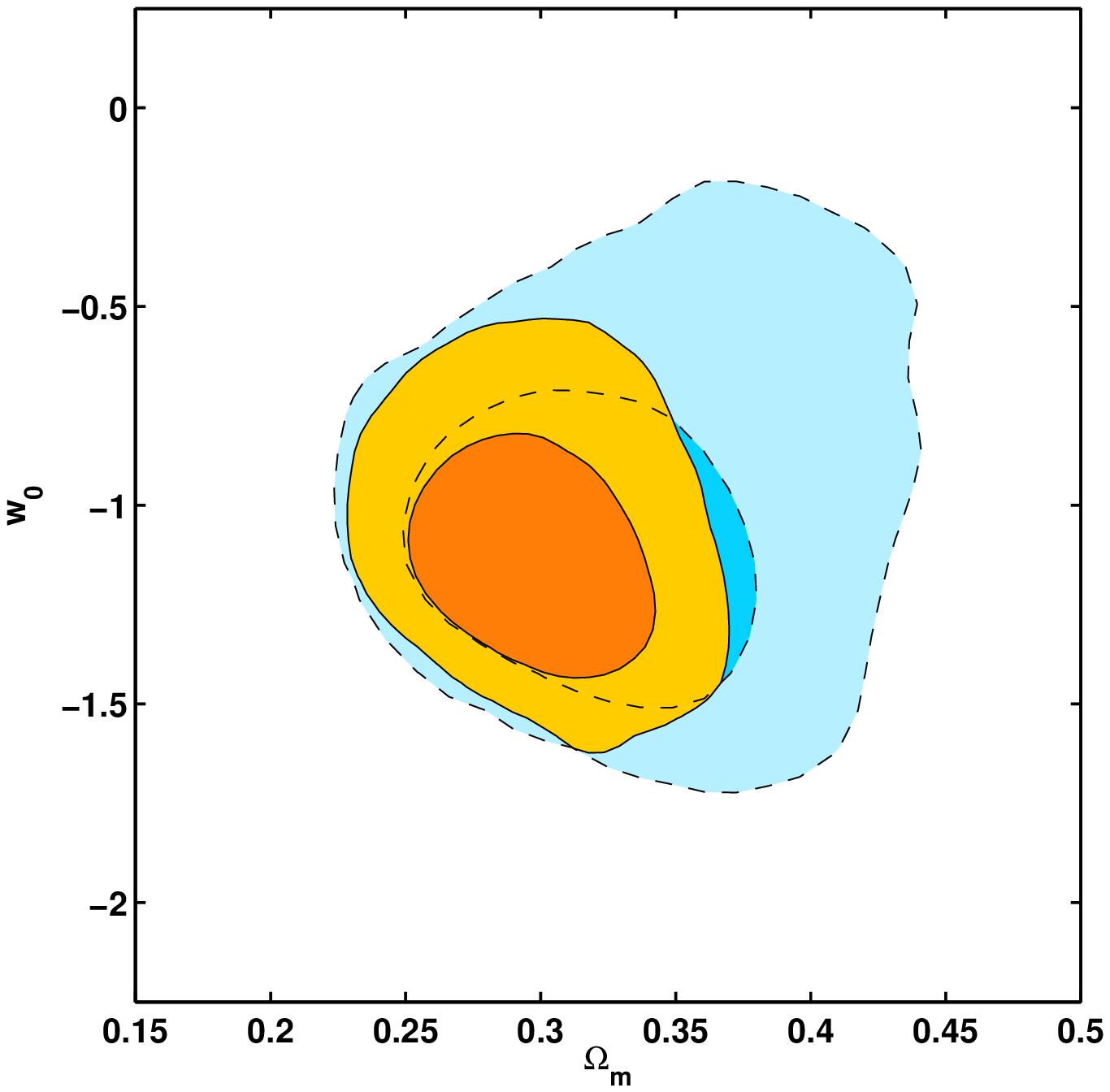}
\caption{The 68.3 and 95.4 per cent confidence limits in the
  $(\Omega_{\rm m},w_{\rm 0})$ planes from the analysis of all three
  data sets combined, assuming either a flat prior (solid contours) or
  including the curvature as an additional free parameter (dashed
  contours). The left panel shows the results for constant $w$ and the
  right panel for a the $z_{\rm t}=1$ dark energy model.}
\label{non}
\end{figure*} 

Using our most general dark energy model, we find that the combination
of 2dF+SNIa+CMB data leads to slightly weaker and more negative values
for $w_{\rm 0}$ and larger values for $w_{\rm et}$, which rule out the
cosmological constant at 1$\sigma$. However, as discussed above, the
departure from the cosmological constant becomes insignificant when we
also include the cluster gas fraction data.

\section{Discussion}
\n{discussion}

In this paper, we have examined the evolution of dark energy using a
combination of three cosmological data sets: CMB anisotropies, SNIa
and the gas fraction in X-ray luminous galaxy clusters. Employing a
minimum of prior information, we have obtained tight constraints on
various key parameters including the dark energy equation of state.
Assuming that the equation of state remains constant with time, and
assuming a flat prior, we measure $w_{\rm 0}=-1.05^{+0.10}_{-0.12}$.
Employing a series of more general dark energy models, we find no
significant evidence for evolution in the equation of state. A
cosmological constant is a good description of the current data. For
models other than a cosmological constant, we have included the
effects of perturbations in the dark energy component to avoid
violating energy--momentum conservation. We have shown that
neglecting perturbations can lead to spuriously tight constraints on
dark energy models, especially for the $w_{\rm et}$ parameter (by up
to a factor two for our most general model).

Although each of the data sets used here probes certain aspects of
dark energy by itself, much tighter constraints are obtained when the
data are combined. The SNIa and cluster data provide the primary
source of information on the evolution of dark energy. Of these, the
SNIa data currently contribute the stronger constraints. However, this
power can only be utilised once a tight constraint on $\Omega_{\rm
m}$, in this case provided by the combination of the cluster+CMB (or
cluster+BBN+HST) data, is included. Using the SNIa data alone, one is
hampered by a degeneracy between the equation of state $w$ and
$\Omega_{\rm m}$ e.g. \cite{Riess:04} (see also \cite{Wang:04}). This
is the reason that some authors have employed a strong prior on
$\Omega_{\rm m}$ in their studies using SNIa data
\citep{alam:04,jassal:04,jonsson:04, jain:04}; without such a prior,
the SNIa data alone cannot provide tight constraints on even constant
equation of state models, much less models with evolution in $w$.
Rather than introducing strong priors, our approach has been to use a
combination of data sets that are complementary in nature and which
allow certain key parameter degeneracies to be broken.

One of the main parameter degeneracies highlighted in previous studies
\citep{Corasaniti:04a} is between the scalar spectral index $n_s$, the
physical baryon density $\Omega_{\rm b}h^2$ and the optical depth to
reionisation, $\tau$; this degeneracy impinges on the measured dark
energy parameters. As noted by \cite{Corasaniti:04a}, the
integrated Sachs-Wolfe effect in the case of an evolving dark energy
equation of state increases the importance of this degeneracy with
respect to constant $w$ models. The left panel of Fig~\ref{degen}
shows this degeneracy for the case of the SNIa+CMB data, using the
$z_{\rm t}=0.35$ dark energy model. The right panel shows how the
degeneracy is lessened when the clusters+CMB data are used. The
combination of clusters+CMB data also leads to a tight constraint on
$H_0$ (e.g. \cite{Allen:03}).

\begin{figure}
\includegraphics[width=3.2in]{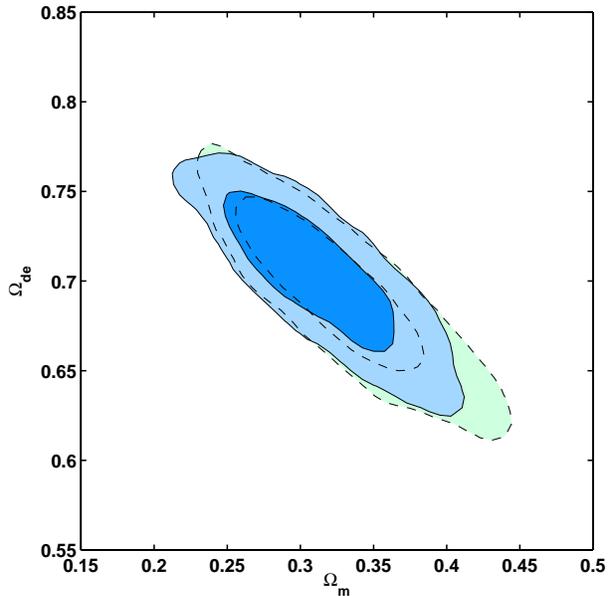}
\caption{The 68.3 and 95.4 per cent confidence limits in the
  $(\Omega_{\rm m}$,$\Omega_{\rm de})$ plane from the analysis of the
  combined cluster+SNIa+CMB data set, with $\Omega_k$ included as a
  free parameter. The solid contours show the constraints for constant
  $w$. The dashed contours show the results for the $z_{\rm t}=1$ dark
  energy model.}
\label{non2}
\end{figure}

The combination of clusters+CMB data even allows us to relax the
assumption that the Universe is flat, although we note that the
computation of MCMC chains in the non-flat case is time consuming when
one wishes to ensure convergence with $R-1\lt 0.1$. (For this reason,
we have only carried out a limited exploration of non-flat models
here.)  In order to avoid unphysical regions of the parameter space
when using non-flat models, we have also included a prior on the
optical depth to reionisation, $\tau<0.3$, in a similar manner to WMAP
team \citep{Spergel03} and \cite{Corasaniti:04a}. Figure~\ref{non}
shows the 68.3 and 95.4 per cent confidence limits in the
($\Omega_{\rm m}$,$w_{\rm 0}$) plane obtained assuming a constant dark
energy equation of state (left panel) and using the $z_{\rm t}=1$
model (right panel), with (dashed lines) and without (solid lines) the
curvature included as a free parameter. Comparison of the curves shows
that the uncertainties in the parameters are increased when the
assumption of flatness is dropped. However, we still have clear
evidence that $w_{\rm 0} \lt -1/3$ and therefore that the Universe is
accelerating at late times. For the constant equation of state model,
we obtain tight constraints on $\Omega_{\rm
k}=-0.017^{+0.020}_{-0.021}$, $\Omega_{\rm m}=0.314^{+0.040}_{-0.036}$
and $\Omega_{\rm de}=0.703^{+0.026}_{-0.030}$. The median parameter
values and parameter values at the peaks of the marginalised
probability distributions (and 68 per cent confidence intervals) for
$w_{\rm 0}$, $w_{\rm et}$ and $\Omega_{\rm m}$ using the non-flat
models are summarised in Table~\ref{tab2}. Figure~\ref{non2} shows the
results in the ($\Omega_{\rm m}$,$\Omega_{\rm de}$) plane.

Finally, it is encouraging to recognise the prospects for advances in
this work over the next $1-2$ years. Further Chandra observations of
X-ray luminous, high-redshift, dynamically relaxed clusters should
lead to rapid improvements in the constraints from the X-ray method.
Continual progress in SNIa studies is expected over the next few years
and the forthcoming, second release of WMAP data should, at the very
least, provide an important, overall tightening of the constraints. In
the long term, the combination of complementary constraints from
missions such as Constellation-X, SNAP and Planck, combining high
precision with a tight control of systematic uncertainties, offers our
best prospect for understanding the nature of dark energy.

\section*{Acknowledgement}
We acknowledge helpful discussions with A.~Lewis, S.~Bridle and
P.~S.~Corasaniti and technical support from R.~M.~Johnstone and
S.~Rankin. The computational analysis was carried out using the
Cambridge X-ray group Linux cluster and the UK National Cosmology
Supercomputer Center funded by SGI, Intel, HEFCE and PPARC. DR is
funded by an EARA Marie Curie Training Site Fellowship under the
contract HPMT-CT-2000-00132. DR thanks KIPAC, SLAC and Stanford
University for hospitality during his visit there. SWA thanks the
Royal Society for support at Cambridge. This work was supported by the
NASA grant *** and the DOE at Stanford and SLAC. JW is supported
by the DOE and the NASA grant NAG 5-10842 at Fermilab.

%\bibliography{/home/weller/biblist}
%\bibliography{/home/jw249/biblist}
%\bibliographystyle{mn2e}

\label{lastpage}
\end{document}